\documentclass[12pt,preprint]{aastex}
\usepackage[numberedappendix]{emulateapj5}
\shorttitle{SEYFERT--STARBURST CONNECTION}
\shortauthors{LAINE ET AL.}
\begin{document}

\submitted{ACCEPTED FOR PUBLICATION IN THE ASTRONOMICAL JOURNAL}
\title{EXAMINING THE SEYFERT -- STARBURST CONNECTION WITH \\
ARCSECOND RESOLUTION RADIO CONTINUUM OBSERVATIONS}

\author{Seppo Laine}
\affil{{\it Spitzer} Science Center, Mail Code 220-6, California Institute of Technology,  
Pasadena, CA 91125}
\email{seppo@ipac.caltech.edu}

\author{Jari K. Kotilainen}
\affil{Tuorla Observatory, University of Turku, 
V\"{a}is\"{a}l\"{a}ntie 20, 21500 Piikki\"{o}, Finland}

\author{Juha Reunanen}
\affil{University of Leiden, Department of Astronomy, PO Box 9513, 2300 
RA Leiden, The Netherlands}

\author{Stuart D. Ryder}
\affil{Anglo-Australian Observatory, PO Box 296, Epping, NSW 1710, 
Australia}

\and

\author{Rainer Beck}
\affil{Max Planck Institut f\"{u}r Radioastronomie, Auf dem H\"{u}gel 69, 
53121 Bonn, Germany}
    
\begin{abstract} 

We compare the arcsecond-scale circumnuclear radio continuum properties between
five Seyfert and five starburst galaxies, concentrating on the search for any
structures that could imply a spatial or causal connection between the nuclear
activity and a circumnuclear starburst ring. No evidence is found in the radio
emission for a link between the triggering or feeding of nuclear activity and
the properties of circumnuclear star formation. Conversely, there is no clear
evidence of nuclear outflows or jets triggering activity in the circumnuclear
rings of star formation. Interestingly, the difference in the angle between the
apparent orientation of the most elongated radio emission and the orientation
of the major axis of the galaxy is on average larger in Seyferts than in
starburst galaxies, and Seyferts appear to have a larger physical size scale of
the circumnuclear radio continuum emission. The concentration, asymmetry, and
clumpiness parameters of radio continuum emission in Seyferts and starbursts
are comparable, as are the radial profiles of radio continuum and near-infrared
line emission. The circumnuclear star formation and supernova rates do not
depend on the level of nuclear activity. The radio emission usually traces the
near-infrared Br$\gamma$ and H$_{2}$ 1--0 S(1) line emission on large spatial
scales, but locally their distributions are different, most likely because of
the effects of varying local magnetic fields and dust absorption and scattering. 

\end{abstract}


\section{INTRODUCTION}
\label{s:intro}

Up to 10\% of nearby galaxies have energetic, non-stellar Seyfert activity  in
their nuclei \citep*[e.g.,][]{ho97a}. Currently it is unknown why this fraction
is 10\%, and not 100\%. One of the most appealing explanations for this puzzle
is the duty cycle hypothesis \citep*[e.g.,][]{era95}. It can be speculated that
{\it all} spiral galaxies (Seyfert activity appears almost exclusively in
spiral galaxies) become active for a limited period of time in their lives.
Since a typical age for a nearby spiral galaxy is 10$^{10}$~years, and 10\% of
nearby spirals have Seyfert nuclei, the activity may last up to 10$^{9}$~years.

The central problem in feeding AGN activity is the supply of adequate fuel.
While tidal disruption of stars in the vicinity of supermassive nuclear black 
holes is unlikely to be efficient enough to supply several
M$_{\odot}$yr$^{-1}$, as is required for feeding quasar activity, the mass
consumption rates in Seyferts  are only on the order of
0.1~M$_{\odot}$yr$^{-1}$ \citep[e.g.,][]{bian03}. Tidal disruption of stars
may be a sufficient mechanism for maintaining LINER activity \citep{era95}, but
the only viable mechanism for feeding Seyfert activity appears to be the
conversion of gas mass into radiation in accretion disks \citep*{shlos90} that
have been speculated to surround supermassive nuclear black holes. In addition
to the possibility of nuclear bars, or ``bars within bars'', magnetic braking
in a circumnuclear ring can drive gas inflow towards a nucleus and its
accretion disk \citep{beck99,beck05}.

Interstellar gas in the central region of a galaxy can come from at least two
sources. First of these is mass loss from evolved stars. This mechanism may 
produce a few 0.1 M$_{\odot}$yr$^{-1}$ within the central few kpc
\citep[e.g.,][]{fab76}. Another source of gas is from secular inflow,
especially in a barred potential. Gravitational torques from bars are known to
be capable of  extracting angular momentum from the gas that piles up along the
leading edge  of the bar, and cause a secular inflow of gas towards the central
region \citep*[e.g.,][]{sim80,nog88,bar91}. If a dynamical resonance (the
so-called inner Lindblad resonance, hereafter ILR; e.g., \citeauthor{con75}
\citeyear{con75}; \citeauthor{ath92} \citeyear{ath92}) exists, gas will tend to
pile up in a ring around the ILR radius (there can be 0, 1, or 2 ILRs, or in
the case of a large central mass concentration, even a third ILR; gas tends to
pile up and form a ring between the two outer ILRs, e.g., \citeauthor{kna95}
\citeyear{kna95}). Properties of such rings have been discussed in numerous
papers. A good and comprehensive summary is given by \citet{buta96}.

The nuclear rings are often locations of vigorous star formation (SF). While
the exact fraction of gas that is consumed by SF in these rings is not known
yet (although it is likely to vary from galaxy to galaxy), it is clear that the
circumnuclear rings play a crucial role in regulating the gas supply to the
nucleus \citep*[e.g.,][]{heller96,regan99}. Therefore, it is of great
importance to determine whether some properties of the circumnuclear rings,
such as a special location of star forming regions, star forming rates,
closeness to the nucleus, etc., are related to the activity class of the
nucleus. Conversely, nuclear activity can have an impact on the properties of 
the starburst ring, e.g., in the form of star-forming activity in the
circumnuclear ring triggered by a radio jet, as demonstrated on large-scale  in
Minkowski's Object \citep[e.g.,][]{fragile04}. These issues can be studied by
looking for similarities and/or differences in the ring properties and the
strength of nuclear activity between Seyfert and starburst galaxies. Earlier
studies of circumnuclear rings have largely concentrated on the use of
ultraviolet, optical, or near-infrared (NIR) images and spectra to study the
properties of SF in these regions
\citep*[e.g.,][]{ben93,bar95,kot96a,kot96b,elm97,elm98,buta99,elm99,buta00,kot00,
per00,reu00,kot01,maoz01,alo01,ryder01}. Direct observations of the molecular
gas also exist \citep[e.g.,][]{saka95,bene96}, mostly via CO that can be used
as a tracer of the more abundant molecular hydrogen with certain caveats, such
as the uncertainty in the conversion of the CO luminosities into total
molecular gas masses, but the spatial resolution in most CO studies has been
insufficient to resolve the rings adequately.

An alternative and potentially more powerful way to study the properties of 
the circumnuclear rings is provided by high resolution radio continuum 
observations. Radio wavelengths are free of extinction effects that dust has in
optical and, for high extinction, even in near-infrared images. The radio
emission coming from the rings is believed to consist of two main components.
First, thermal emission, tracing bremsstrahlung from free electrons in
\ion{H}{2} regions, thermalized by the  optically thick medium. Second,
nonthermal emission, coming from electrons spiraling in magnetic fields, either
surrounding recent supernova remnants (seen in high resolution maps), or in the
general magnetic field of the underlying  disk galaxy (seen in low resolution,
high sensitivity maps, e.g.,  \citeauthor{tur94} \citeyear{tur94}), or in the 
enhanced magnetic field of the ring \citep{beck05}. Emission from AGN is also
nonthermal. Nonthermal emission usually dominates the emission maps seen at 6
and 20 cm, whereas at shorter wavelengths the thermal fraction can be
substantial or even dominate \citep[e.g.,][]{tur94}. 

An interesting feature of radio continuum maps of several active galaxies is
the strong point source in the nucleus, and jets or highly collimated outflows
emanating from the nucleus. In Seyfert galaxies, these jets are usually  seen
at a few hundred pc scale \citep{schmitt01}, but they can extend to kpc scales
(e.g., NGC~4258: \citeauthor{van82} \citeyear{van82}; NGC~7479:
\citeauthor{lai98} \citeyear{lai98}). The orientation of the jets with respect
to the rotation axis or disk plane of the galaxy has been observed to be fairly
random \citep{schmitt97}. Such a collimated outflow has the potential of
triggering SF in a circumnuclear region as hinted by, e.g.,
\citet{bra98} and Minkowski's Object \citep{fragile04}. The recent study by \citet{knapen05} suggests that galaxies with
an active nucleus may have rings more often than galaxies without such a
nucleus. One of the main objectives of the current paper is to investigate if
evidence for jets and triggering of SF exists among nearby Seyfert
galaxies, and to contrast their circumnuclear properties to those of nearby
starburst galaxies. The triggering of SF in starburst galaxies is
most likely external to the circumnuclear region, as there is no strong nuclear
outflow component.

Several previous investigations of the radio continuum morphology in the
central regions of nearby disk galaxies have been performed
\citep[e.g.,][]{colli94,sai94,tur94,bra98,for98,mor99,beck00,nag00,the00,ho01,beck05}.
However, none of these earlier investigations either had a balanced sample
between Seyfert and non-Seyfert galaxies, or supporting near-infrared line
images of 2.166 $\mu$m Br$\gamma$ or 2.122 $\mu$m H$_{2}$ 1--0 S(1) emission to
study the star forming morphology and properties. Many of the previously listed
studies also had resolution and sensitivity combinations that were not ideally
suited to detecting extended but resolved circumnuclear structure, for which a
resolution of order 1--2 arcseconds is needed, together with a sensitivity to
structures tens of arcseconds in extent. Our new observations are ideal for the
study of such structure. In addition, we have gathered data at several
wavelengths, approximately matched in spatial resolution, to further improve
the sensitivity to emission from different mechanisms at various spatial scales
and to provide some information on the radio spectral indices at various
locations in our targets.

\section{SAMPLE AND OBSERVATIONS}
\label{s:sampleobs}

Our sample consists of both Seyfert and starburst galaxies, all of which are
either barred or merging systems, thus making recent gas inflow likely. We
include all the nearby (within 100 Mpc) Seyfert, starburst or
merging/interacting galaxies which have NIR line images of either 2.166 $\mu$m 
Br$\gamma$ or 2.122 $\mu$m H$_{2}$ 1--0 S(1) emission or both available, and
which do not have existing sensitive radio continuum observations at 1--2
arcsec resolution at 3.5, 6 or 20 cm. VLA sensitivity in the L-band (20 cm)
improved by a factor of two in the early 1990s when the receivers were upgraded
(P. Lilly and R. Perley, private communication). The C-band receivers have 
also improved, resulting in a sensitivity improvement, although it is not as 
dramatic as in the L-band. Therefore, in some cases we reobserved galaxies
that  had no VLA radio continuum observations at a comparable resolution since
late 1980s. 

General parameters of the ten galaxies that satisfied our sample selection
criteria are listed in Table~\ref{tab1}. Of our sample galaxies, five have
Seyfert classifications (one Seyfert 1 and four Seyfert 2s) and five have
starburst nuclei, and at least two are in merging or interacting systems. A
sample of this size is useful in a pilot study such as ours to look
for any trends and differences between the various nuclear activity classes. We
do not distinguish between the  various activity subclasses such as Seyfert 1
and Seyfert 2 galaxies in the  following, but only contrast the properties of
Seyfert galaxies against non-Seyfert galaxies.

\begin{table*}
\caption{GENERAL PARAMETERS OF THE GALAXY SAMPLE. \label{tab1}}
\begin{tabular}{lcccccccc}
\tableline
\tableline
 Galaxy & Hubble &  Activity &  V$_{hel}$ & D &  $B_{\rm T}$ & Bar Length &
Deproj. Bar & NIR Data\\
&  Type &  Type & (km~s$^{-1}$) & (Mpc) & (mag) & (kpc) & Ellipticity &
Available\\ 
\tableline
NGC~520  & S pec & Starburst (Merger) & 2281 & 27.8 & 12.2 & \nodata
& \nodata & H$_{2}$, Br$\gamma$\\
IC 342   & SAB(rs)cd & Starburst & 31 & 3.9 & 9.1 & 1.7 &
0.43 & H$_{2}$, Br$\gamma$\\
NGC~4536 & SAB(rs)bc & Starburst & 1808 & 13.3 & 11.2 &
1.2 & 0.37 & H$_{2}$\\
NGC~6240$^{a}$ & I0: pec & Sy2 (Merger) & 7339 & 96.1 & 13.8 &
\nodata & \nodata & H$_{2}$\\
NGC~6574$^{a}$ & SAB(rs)bc & Sy2 & 2282 & 35.0 & 12.8 & 1.7 &
0.47 &  H$_{2}$, Br$\gamma$\\
NGC~6764$^{a}$ & SB(s)bc & Sy2 & 2416 & 37.0 & 12.6 & 9.9 &
0.69 &  H$_{2}$, Br$\gamma$\\
NGC~7469 & (R')SAB(rs)a & Sy1.2 & 4892 & 64.4 & 13.0 &
0.7 & 0.52 & Br$\gamma$\\
NGC~7479$^{a}$ & SB(s)c & Sy2 & 2381 & 32.4 & 11.6 & 7.5 &
0.74 & H$_{2}$, Br$\gamma$\\
NGC~7714$^{a}$ & SB(s)b:pec & Starburst & 2798 & 36.9 & 13.0 & 7.7 & 0.52 & H$_{2}$, Br$\gamma$\\
NGC~7771$^{a}$ & SB(s)a & Starburst & 4277 & 56.4 & 13.1 &
8.8 & 0.62 & H$_{2}$, Br$\gamma$\\
\tableline
\end{tabular}
$^{a}$Galaxy also observed at 3.5 or 6 cm.
\tablecomments{Galaxy names (col. 1), Hubble type from NED (col. 2), activity
or interaction class (col. 3), heliocentric velocity from NED (col. 4),
distance from \citeauthor{tul88} \citeyear{tul88} for galaxies closer than 40 Mpc, and from Hubble law
with H$_{0}$ = 75 km~s$^{-1}$~Mpc$^{-1}$ and $q_{\rm 0}$ = 0.5 (col. 5),
$B_{\rm T}$ magnitude from NED (col. 6), deprojected bar length in kpc (col.
7), deprojected bar ellipticity (col. 8), available NIR data (col 9). The bar
information was derived or taken from 2MASS $H$ or $K$ images (IC~342, NGC~4536),
\citeauthor{kot00} \citeyear{kot00} (NGC~6574), \citeauthor{kot01}
\citeyear{kot01} (NGC~7714), \citeauthor{reu00} \citeyear{reu00} (NGC~7771), 
\citeauthor{lai02} \citeyear{lai02} (NGC~7469, NGC~7479), and J. Kotilainen
et al., unpublished (NGC~6764).}

\end{table*}

\begin{figure*}
\centering
\includegraphics[width=6in]{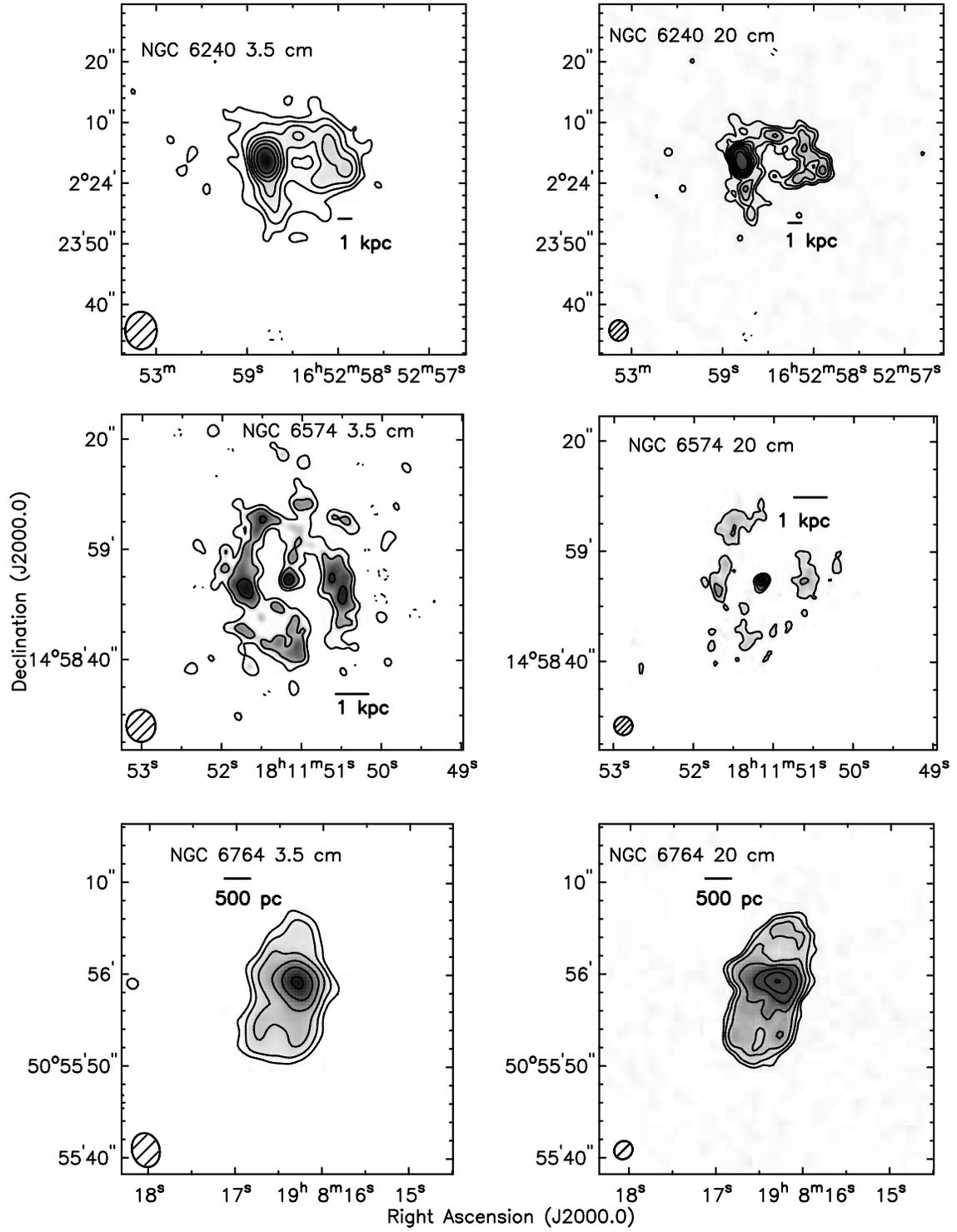}
\caption{Gray-scale and contour images of the radio continuum 
emission in three sample galaxies. The observed wavelength is given within the
images. The contours are at (-8 -4 4 8 16 32 64 128 etc.) 
times the one sigma noise level. Any negative areas are surrounded by dashed
lines. The sigma (rms) values are given in Table~\ref{tbl3}. The beam size is marked with the ellipse on the 
lower left corner of each image. Size scale is shown with a horizontal bar.\label{fig1}}
\end{figure*}

\setcounter{figure}{0}

\begin{figure*}[th]
\centering
\includegraphics[width=6in]{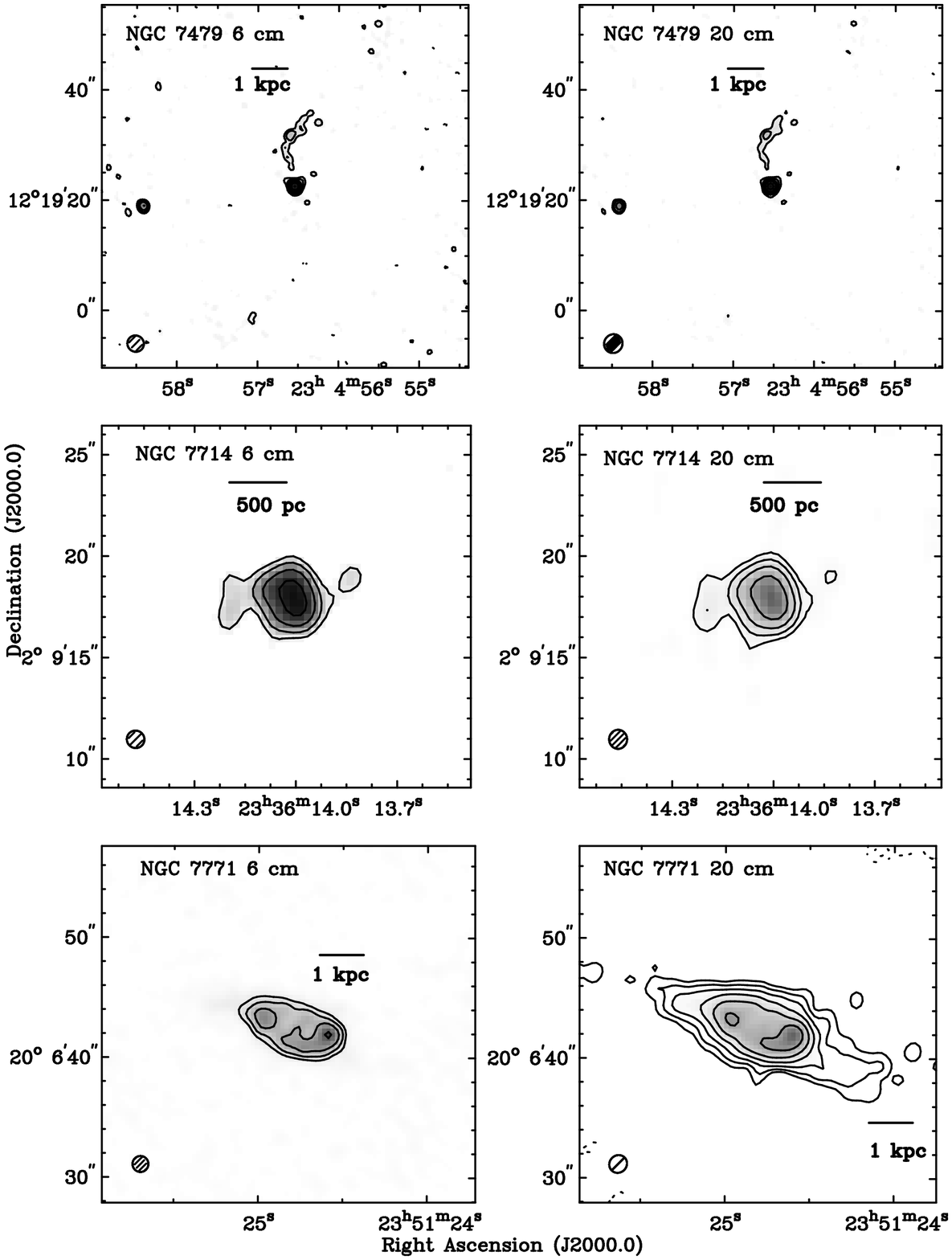}
\caption{Continued.}
\end{figure*}

Very Large Array (VLA) observations of the sample galaxies were taken during
three separate observing sessions (see Table~\ref{tab2}). The galaxies were
mixed both among and within the two observing sessions to maximize the $u-v$
coverage. Weather  conditions were excellent for the 6 cm run, while cloud
cover was between 15\% and 75\% for the 3.5 and 20 cm runs. These wavelengths
were selected because they give approximately comparable spatial resolutions
when observed with the VLA in its A configuration (20 cm), B configuration (6
cm), and C configuration (3.5 cm). The 3.5 cm band was used instead of the 2 cm
band because of its higher sensitivity. The large range of wavelengths also
gives us some leverage to  inspect differences in the distribution of thermal
(3.5 cm) and non-thermal (20 cm) emission. The absolute flux calibration is
based on 5--10 minute scans of 3C 48 and 3C 286. The target galaxy observations
were bracketed between 1--2 minute scans of nearby phase calibrators, and their
fluxes were bootstrapped to those of the primary calibrators. The derived
fluxes for the phase calibrators were then compared to values found in the VLA
calibrator flux density database to check their reasonableness. Uncertainties
in the quoted fluxes are dominated by the uncertainty in setting the absolute
flux scale, and are estimated to be less than 10\%. Uncertainties in positions
are estimated to be less than 0\farcs 1. 

\setcounter{figure}{0}

\begin{figure*}[th]
\centering
\includegraphics[width=6in]{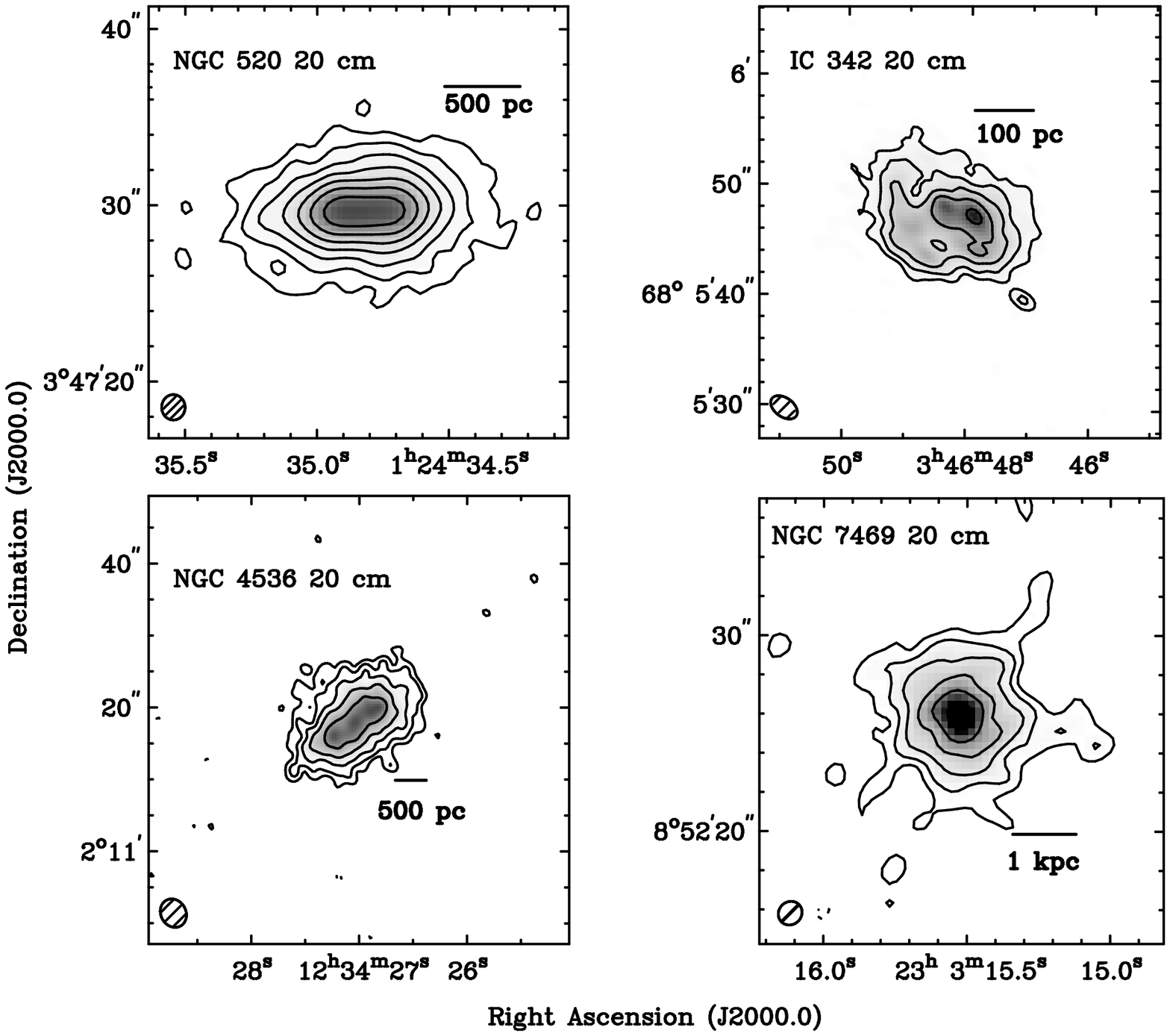}
\caption{Continued.}
\end{figure*}

For maximum sensitivity we used both intermediate frequencies, with bandwidths
of 50 MHz, and separated by 50 MHz. After minimal editing of the visibilities
and calibration, we used the AIPS task IMAGR to Fourier transform the observed
visibilities into brightness distribution maps on the sky. The ``dirty'' maps
were deconvolved using the CLEAN procedure \citep{hog74,cla80} by  placing
``clean boxes'' around the central area of the map (the galaxy) and around
nearby background sources that may have sidelobes that cause additional noise
in the mapped field. The images were cleaned down to about three times the
theoretical root mean square (rms) noise limit of the maps. We used no tapering
and set the ROBUST parameter in IMAGR to zero, resulting in maps which are a
compromise between maximum resolution and maximum sensitivity. Our typical
synthesized beam sizes, tabulated in Table~\ref{tbl3}, have FWHM values around
1\farcs 3 at 6 and 20 cm, and 2\farcs 5 at 3.5 cm. We applied phase-only
self-calibration for the brightest sources in the sample, resulting in a slight
improvement in the S/N ratio. The expected theoretical rms sensitivities were
obtained in most cases where sidelobe contamination from background sources is
not severe. 

\begin{table*}
\caption{PARAMETERS OF THE VLA OBSERVING RUNS. \label{tab2}}
\begin{tabular}{lccc}
\tableline
\tableline
\multicolumn{1}{c}{Parameter} &  Session 1 &  Session 2 & Session 3 \\
\tableline
Dates\dotfill & 1999 Jul 25, 27 & 1999 Nov 26 & 2000 Jun 3--4 \\
Configuration\dotfill & A & B & C \\
Maximum number of antennas\dotfill & 27 & 26 & 27 \\
Length of observations (hr)\dotfill & 2.5, 4 & 3 & 2 \\
Frequency (GHz)\dotfill & 1.4250 & 4.8601 & 8.4601 \\
Bandwidth (MHz)\dotfill & 100 & 100 & 100 \\
Primary flux calibrator\dotfill & 3C 286, 3C 48 & 3C 48 & 3C 286 \\
Assumed flux of primary flux calibrator (Jy)\dotfill & 14.554, 16.319 & 5.516 & 5.181 \\
\tableline
\end{tabular}
\end{table*}

\section{RESULTS}
\label{s:images}

\subsection{Radio Structure and Power}

\begin{figure*}[th]
\centering
\includegraphics[width=4in]{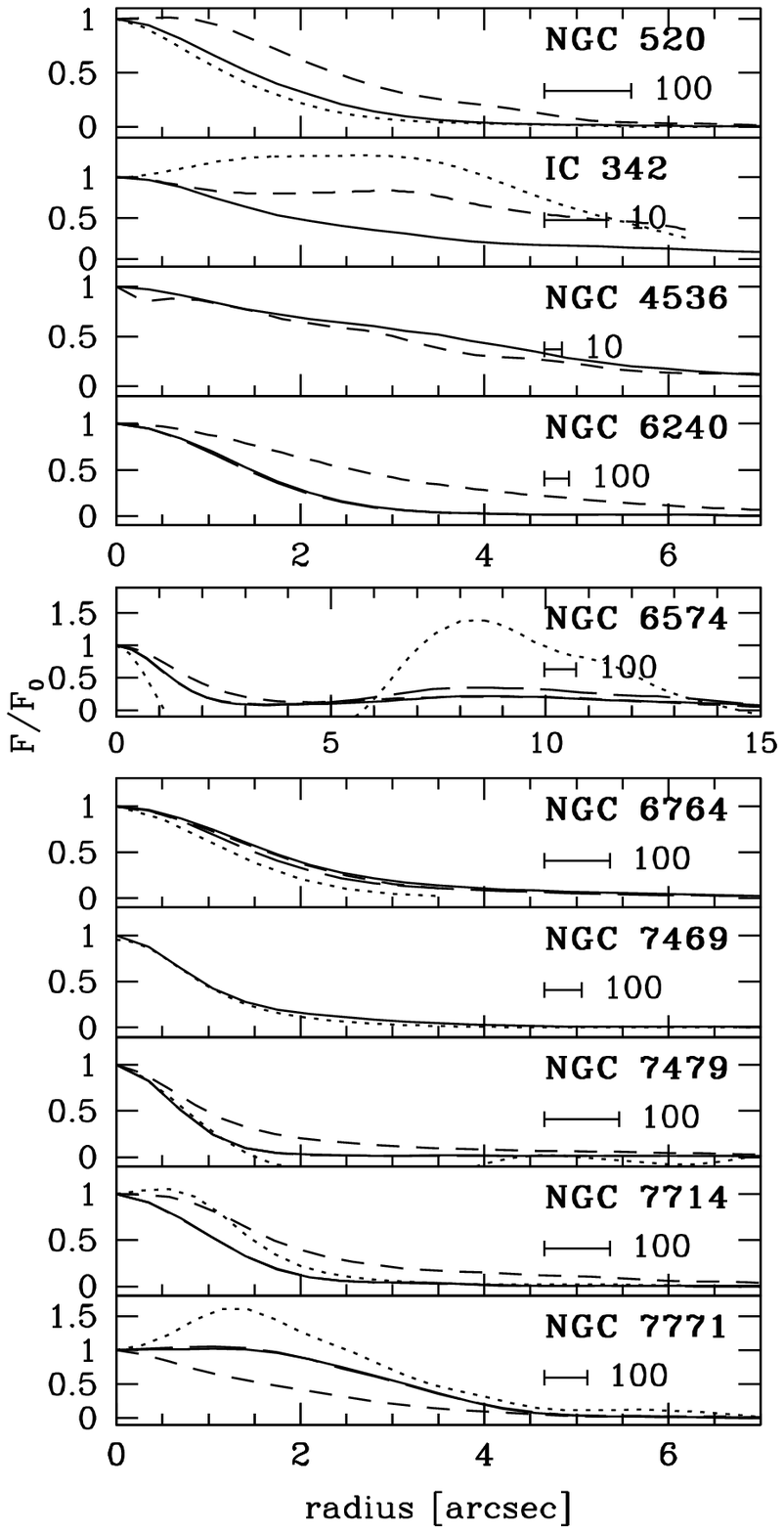}
\caption{Radial surface brightness profiles for the ten sample galaxies.
The 20 cm radio continuum profile is shown with a solid line, the 3.5/6 cm
profiles with a long-dash line, the H$_{2}$
1--0 S(1) emission with a short-dash line, and the Br$\gamma$ emission profile
with a dotted line. The profiles are normalized to the same relative intensity
close to the center. The scale in parsecs (10 or 100) is also indicated.\label{profilesfig}}
\end{figure*}

Gray-scale and contour images of the radio continuum emission at 3.5, 6, and 20
cm from our sample galaxies are presented in Figure~\ref{fig1}. The integration
times and several other parameters of the final maps are given in
Table~\ref{tbl3}. Figure~\ref{profilesfig} shows the radial surface brightness
profiles of 3.5, 6, and 20 cm radio continuum, H$_{2}$ 1--0 S(1) emission, and
Br$\gamma$ emission. In Figures~\ref{n520fig}--\ref{n7771fig} we show the radio
continuum emission together with either molecular hydrogen H$_{2}$ 1--0 S(1) or
Br$\gamma$ emission maps, or both. 

\begin{table*}
\caption{OBSERVATIONAL AND MEASURED PARAMETERS. \label{tbl3}}
{\scriptsize
\begin{tabular}{lcccccccccc}
\tableline
\tableline
 Galaxy & \multicolumn{5}{c}{Map Parameters} & 
\multicolumn{5}{c}{Source Parameters} \\
  &  $\lambda$ & Int. Time & Beam &  P. A. & 
 rms &  Total Flux & Total Flux & Log Total Radio &  U/E$^{a}$
&  Morph. \\
  &  (cm) & (sec) & (arcsec $\times$ arcsec) &  (deg) &
 ($\mu$Jy~bm$^{-1}$) &  (VLA; Jy) & (Single dish; Jy) &
 Power (W Hz$^{-1}$) & &   \\
\tableline
NGC~520 & 20 & 700 & 1.44 $\times$ 1.33 & 0 & 65 & 0.170 & 0.158 & 22.2 & 0\% & C \\
IC 342 & 20 & 810 & 1.70 $\times$ 1.07 & 54 & 90 & 0.131 & 2.25 & 20.2 & 0\%& R, E \\
NGC~4536 & 20 & 660 & 1.67 $\times$ 1.42 & 29 & 51 & 0.131 & 0.126 & 21.4 & 0\% & L, E \\
NGC~6240 & 20 & 760 & 1.48 $\times$ 1.33 & 0 & 43 & 0.108 & 0.59 & 23.1 & 29\% & C, R\\
NGC~6240 & 3.5 & 700 & 2.72 $\times$ 2.28 & 0 & 66 & 0.115 & ... & 23.1 & 40\% & C, R \\
NGC~6574 & 20 & 3590 & 1.34 $\times$ 1.32 & 0 & 42 & 0.030 & ... & 21.6 & 4\% & C, R \\
NGC~6574 & 3.5 & 1940 & 2.33 $\times$ 2.11 & 0 & 29 & 0.026 & ... & 21.6 & 2\% & C, R\\
NGC~6764 & 20 & 780 & 1.39 $\times$ 1.26 & 135 & 53 & 0.101 & ... & 22.2 & 12\% & C, L \\
NGC~6764 & 3.5 & 1040 & 2.50 $\times$ 2.00 & 16 & 54 & 0.029 & ... & 21.7 & 24\% & C, L \\
NGC~7469 & 20 & 1080 & 1.40 $\times$ 1.29 & -46 & 211 & 0.181 & 0.192 & 23.0 & 18\% & C, E \\
NGC~7479 & 20 & 610 & 1.38 $\times$ 1.32 & -32 & 29 & 0.005 & 0.100 & 20.8 & 50\% & C, L\\
NGC~7479 & 6 & 1590 & 1.21 $\times$ 1.17 & 56 & 21 & 0.006 & 0.041 & 20.9 & 39\% & C, L \\
NGC~7714 & 20 & 4770 & 1.38 $\times$ 1.30 & -2 & 83 & 0.0153 & 0.047 & 21.4 & 29\% & C, E \\
NGC~7714 & 6 & 4340 & 1.28 $\times$ 1.26 & 58 & 95 & 0.0141 & ... & 21.4 & 30\% & C, E\\
NGC~7771 & 20 & 1040 & 1.33 $\times$ 1.24 & -45 & 26 & 0.0252 & 0.128 & 22.0 & 6\% & R, E\\
NGC~7771 & 6 & 2270 & 1.16 $\times$ 1.15 & 0 & 21 & 0.0263 & 0.047 & 22.0 & 8\% & R, E \\
\tableline
\end{tabular}
}
$^{a}$Ratio of unresolved to resolved emission flux densities. The nuclear
source was fitted with a Gaussian and if its FWHM values were practically
identical to the FWHM values of the beam, the nuclear source was called
``unresolved.''
\tablecomments{The uncertainties in the total fluxes are between 1 and 6 mJy. 
The uncertainties in the radio powers are less than 0.1 W Hz$^{-1}$.
Symbols for the morphology are as follows: C = core, E = extended, L = linear, R = ring.
The references for single dish fluxes are: NGC~520, NGC~4536, NGC~7469, NGC~7479, 
NGC~7714, NGC~7771 \citeauthor{mira} \citeyear{mira} (at 21 cm) ; IC 342 
\citeauthor{baker} \citeyear{baker} (at 21 cm); NGC 6240 \citeauthor{cond88} 
\citeyear{cond88} (at 21 cm); NGC 7771 \citeauthor{sramek} \citeyear{sramek} 
(at 6 cm); NGC 7479 \citeauthor*{cond91} \citeyear{cond91} (at 6 cm).
} 
\end{table*}

The morphology of the radio continuum emission varies from centrally peaked 
(NGC~520, NGC~7469, NGC~7714) to elongated or ring-like emission (IC 342, 
NGC~4536, NGC~6240, NGC~6574, NGC~7771) to outflows or jets (NGC~6764,
NGC~7479). The orientations of the circumnuclear radio continuum structure at
20-cm  and the position angle of the main disk of the underlying galaxy are
given in Table~\ref{tab4}, together with the projected difference between the
two and the inclination angle. The major axis of the radio continuum emission
was determined by fitting a two-dimensional Gaussian to the brightest point,
after smoothing the image to 10 arcsec resolution, and used as the orientation
of the radio emission. Most commonly the detected circumnuclear radio continuum
structures are aligned close to the major axis of the galaxy, implying that
they are most likely emission coming from synchrotron radiation from electrons
spiraling in the magnetic fields of the disk. For NGC~7469 and NGC~7714 the
orientation of the radio continuum emission was very difficult to determine,
and therefore the discrepant values for these galaxies imply that there is only
a more or less circular, extended nuclear component. The discrepant value for
NGC~6240 is due to the peculiar western radio continuum loop, discussed in
detail by \citet*{col94}. In NGC~6764 the almost perpendicular radio continuum
emission suggests an outflow. The outflow will be discussed in more detail in 
another paper (S. Leon et al. 2005, in preparation). The Seyfert systems in our
sample have an average difference  of 60 degrees in the orientation of their
radio continuum emission with respect to the position angle of the galaxy major
axis. On the other hand, for the starburst galaxies the average value for the
difference is 20 degrees, or only 8 degrees if we exclude the uncertain case of
NGC~7714. Such a difference between Seyfert and starburst galaxies could be
explained by the existence of out of plane radio jets and outflows that take
place preferentially in Seyfert galaxies. Our sample is too small to
meaningfully compare the results to those of \citet{schmitt97}, who found that
the nuclear radio continuum  structures in Seyfert~2 galaxies appear to have a
fairly arbitrary orientation with respect to the galaxy major axis.

\begin{table*}
\caption{ORIENTATION AND SIZE SCALE OF STRUCTURES. \label{tab4}}
\begin{tabular}{lccccc}
\tableline
\tableline
 Galaxy & Orientat. of R.C. & Galaxy Position Angle & Difference &
Major Axis Size & Inclination\\
  &  (deg) & (deg) & (deg) &  (kpc) & (deg)\\
\tableline
NGC~520 & 97 & 93 & 4 & 1.6 & 70\\
IC 342 & 63 & 40 & 23 & 0.3 & 31 \\
NGC~4536 & 128 & 130 & 2 & 0.9 & 66 \\
NGC~6240 & 95 & 20 & 75 & 5.1 & 70 \\
NGC~6574 & 10 & 160 & 30 & 4.1 & 37 \\
NGC~6764 & 161 & 62 & 81 & 2.2 & 62 \\
NGC~7469 & 28 & 125 & 83 & 2.9 & 47 \\
NGC~7479 & 170 & 22 & 32 & 2.5 & 51 \\
NGC~7714 & 126 & 4 & 68 & 1.1 & 52 \\
NGC~7771 & 69 & 68 & 1 & 3.6 & 75 \\
\tableline
\end{tabular}

\tablecomments{The galaxy position angle of NGC~520 is deduced only from the inner 
12 arcsec \citep{kot01}. The galaxy position angle for IC 342 is deduced from \ion{H}{1}
and CO observations of \citeauthor*{cros00} \citeyear{cros00} and 
\citeauthor{cros01} \citeyear{cros01}. The galaxy position angles of NGC~4536,
NGC~6240, NGC~6574, NGC~6764, NGC~7469, NGC~7714, and NGC~7771 have
been taken from RC3 \citep{vau91}. The galaxy position angle of NGC~7479 has been taken 
from \citeauthor{lai98} \citeyear{lai98}. Inclination references: NGC~520 \citep{yun01}; IC 342 
\citep{cros00}; NGC~4536 \citep*{jogee05}; NGC~6240 \citep{tac99}; NGC~6574
\citep{saka99}; NGC~6764 \citep{gros85}; NGC~7469 \citep{genz95}; NGC~7479
\citep{lai98}; NGC~7714 \citep{gros85}; NGC~7771 \citep{nord97}.}

\end{table*}

\begin{table*}
\caption{CAS PARAMETERS OF GALAXIES. \label{tab35}}
\begin{tabular}{lccc}
\tableline
\tableline
Galaxy & Asymmetry & Clumpiness & Concentration\\
\tableline
NGC~520 & 0.37 & 0.42 & 0.59\\
IC 342 & 0.56 & 0.25 & 0.39\\
NGC~4536 & 0.38 & 0.21 & 0.20\\
NGC~6240 & 0.90 & 0.38 & 0.73\\
NGC~6574 & 0.50 & 0.39 & 0.11\\
NGC~6764 & 0.66 & 0.28 & 0.55\\
NGC~7469 & 0.27 & 0.30 & 0.69\\
NGC~7479 & 0.58 & 0.57 & 0.56\\
NGC~7714 & 0.27 & 0.47 & 0.81\\
NGC~7771 & 0.43 & 0.31 & 0.38\\
Seyf. Avg. & 0.58$\pm0.21$ & 0.38$\pm$0.10 & 0.53$\pm$0.22 \\
Non-Seyf. Avg. & 0.40$\pm0.09$ & 0.33$\pm$0.10 & 0.47$\pm$0.21 \\
\tableline
\end{tabular}
\tablecomments{The asymmetry for NGC 7469 was calculated by masking out all the
pixels below eight times the sigma noise level in the original image. For 
other galaxies, with more extended emission, four times the sigma noise level
in the original image was used for masking.}
\end{table*}

In general, the radial radio and near-infrared line emission profiles follow
each other (Figure~\ref{profilesfig}). Exceptions are IC~342, where the high
surface brightness near-infrared line emission extends to larger radii than the
radio emission, NGC~6574, where Br$\gamma$ emission has a peak that is not seen
in the other three profiles, at radii around 10 arcsec from the center, and
NGC~7771 where the Br$\gamma$ profile rises from the nucleus within 2 arcsec,
indicating the location of the starburst ring, while the other profiles are
either constant or falling. The differences in the  radio continuum morphology
and near-IR line emission morphology are likely to be due to the varying the
spatial distribution of dust absorption (near-IR lines) and magnetic fields
(radio continuum). The radio profiles for galaxies which had radio observations
at two wavelengths also follow each other well, except for NGC~6574. For
NGC~6574 the starburst ring is better visible at  3.5~cm. No systematic
differences are seen between the profile trends and the relative radial
displacement of the various emissions between Seyfert and starburst galaxies. 

The size scale of the radio continuum emission in kpc is also listed in
Table~\ref{tab4}. This was taken as the major axis length of the FWHM of the
two-dimensional Gaussian fit to the 20-cm emission in an image smoothed to 10
arcsec resolution, except for galaxies which essentially only have a nuclear
component and little extended emission. For these, and NGC~6574 where the
extended emission is well separated from the nuclear emission, the 4-sigma
contours in Figure~\ref{fig1} were used. The average scale size of the radio
continuum emission in Seyferts is 3.4 kpc (rms 1.1 kpc) and in starburst 
galaxies 1.5 kpc (rms 1.1 kpc). However, two of the nearest galaxies are in the
starburst sample, and the lack of information in our observations at large
scales (resolved out by the interferometer) does not warrant us to conclude
that the size scales of  circumnuclear radio continuum emission are different
between the Seyfert and starburst galaxies. However, this is an interesting
result that warrants  follow-up studies using larger samples that are selected
to have galaxies at comparable distances.

In Table~\ref{tab1} we list the deprojected bar lengths and ellipticities for
our sample galaxies when a bar was detected. The deprojection method was
identical to that used by \citet{lai02}. We note that the galaxies with
ring-like morphologies (except the peculiar case of NGC~6240, see below) all
have bars. It has been established quite well that bars create resonance rings
in the circumnuclear area \citep[e.g.,][and references therein]{sel93}. The
connection between the nuclear activity, bar properties, and circumnuclear
rings is less well established. Within our small sample, four galaxies have
long and strong (high ellipticity) bars (NGC~6764, NGC~7479, NGC~7714, and
NGC~7771). Two of these are Seyferts and two are starbursts. Four of our sample
galaxies have short and weak bars (IC 342, NGC~4536, NGC~6574, and NGC~7469).
Again, two of these are Seyferts and two are starbursts. All the tabulated
types of radio continuum morphologies are found among the galaxies with long
and strong  bars (centrally peaked, NGC~7714; linear ``jet-like'', NGC~7479;
ring-like, NGC~7771; and outflow, NGC~6764). Ring, linear, and centrally
peaked  radio continuum morphologies are also found among the galaxies with
short and weak bars. 

The distance-independent radio powers at 20 cm are listed in Table~\ref{tbl3}.
One should note that these powers are measured within widely varying physical
sizes in the galaxies (from $\sim$400 pc in IC~342 to $\sim$9.8 kpc in
NGC~6240, see Table~\ref{tab4}). It appears that the Seyfert galaxies are
slightly more powerful than the starburst galaxies in radio wavelengths. The
merging system NGC~6240 has the largest radio  power.  There also appears to be
a slight tendency towards higher radio powers among the  more strongly barred
galaxies. This is in agreement with the result of \citet{beck02} who found that
the average surface brightness in radio continuum correlates with the relative
bar length. 

Table~\ref{tbl3} also includes an attempt to estimate the fraction of the
radio continuum emission coming from the nuclear unresolved component, after
fitting a Gaussian to it to estimate its flux density. In all cases, the
extended emission dominates. Three out of five non-Seyfert galaxies have no
unresolved component, while the average fraction of flux contributed by the
unresolved component is about 25\% for the Seyfert galaxies.

Finally, Table~\ref{tab35} shows the CAS (Concentration, C, Asymmetry, A, and
Clumpiness, S) measurements at 20 cm for all the sample galaxies. These 
measurements were made in a way similar to that in \citet{conse03}.
Concentration was measured in 3 and 10 arcsec radius  apertures. In cases where
the central structure is clearly elliptical (NGC~520, NGC~4536, NGC~7771) the
measurements were made in elliptical annuli aligned along the major axis of the
structure and having a major axis length of 3 and 10 arcsec. Naturally, since
the distances of the  galaxies vary by a large factor, these apertures measure
different physical scales, but we consider that it is more important to measure
the concentration regardless of the physical scale, since in almost all
galaxies most of the detected radio continuum emission lies in scales 10--20
arcsec or less. The galaxies with the highest concentration indices are NGC~7771,
NGC~6240, and  NGC~7469, as also seen in the radio continuum emission figures.
The asymmetry was calculated by rotating the image by 180 degrees, then
subtracting the rotated image from the original, taking the absolute value of
the difference, and dividing the result by the original image. In addition, we
only considered regions that had emission at higher than 4-sigma level in the
original image (which was thus used as a mask). If a galaxy was perfectly
symmetric, then the asymmetry index would have a value of zero. High values
indicate a large degree of asymmetry. In NGC~7469 we used 8-sigma as the
threshold for masking. The most asymmetric galaxy is NGC~6240 with its
one-sided emission structure. Clumpiness was calculated by subtracting an image
smoothed to 5 arcsec resolution from the original image, then  dividing the
result by the original image. Thus, clumpy galaxies will have large clumpiness
index values. The uncertainties in the measured quantities are estimated to be
$<$~0.15 for asymmetry, $<$~0.1 for clumpiness, and $<$~0.05 for concentration.

The average asymmetry, clumpiness, and concentration are not statistically
different between Seyfert and non-Seyfert galaxies (Table~\ref{tab35}).
Therefore, we do not find any differences in the circumnuclear CAS properties
in our sample of Seyfert and starburst galaxies.	

\subsection{Star Formation and Radio Emission}

We searched the literature for SF and supernova (SN) rates in the central
regions of our sample galaxies. The results are displayed in
Table~\ref{table5}. The given numbers have been adjusted to our adopted
distances as given in Table~\ref{tab1}. Unfortunately, it was not possible to
find the SF and SN information for all the sample galaxies. The SN rate was 
estimated from Br$\gamma$ fluxes together with starburst models, or from radio 
continuum (usually at 20 cm). Since there exists a relationship between
Br$\gamma$ and non-thermal radio continuum fluxes, the two methods are tied to
each other. We expect Type II supernovae to be the main contributors to 
the cosmic rays emitting the observed radio continuum in the circumnuclear 
starburst regions \citep[e.g.,][]{pannuti00}. Note that the nonthermal radio 
continuum emission intensity strongly depends on the strength of the magnetic 
field which may also be enhanced in the starburst regions \citep{beck05}. The
SF rates in our sample galaxies vary between quiescent SF (a few tenths of
M$_{\sun}$~yr$^{-1}$) to several M$_{\sun}$~yr$^{-1}$. The SF rates between
Seyfert and starburst galaxies do not have any obvious differences. It is also
well known that many Seyfert 2 galaxies have associated circumnuclear
starbursts \citep[e.g.,][]{pogge89}. Most of the SN rates are between 0.01 and
0.15~yr$^{-1}$, and there is no obvious difference between Seyferts and
starbursts. The highest SN rate, perhaps not surprisingly, is in the merging
system NGC~6240.

\begin{table*}
\caption{STAR FORMATION AND SUPERNOVA RATES. \label{table5}}
\begin{tabular}{lcccc}
\tableline
\tableline
 Galaxy &  SF Rate &  SN Rate & Reference & SF Rel. RC\\
  &  (M$_{\odot}$~yr$^{-1}$) &  (yr$^{-1}$) &  
&    \\
\tableline
NGC~520 & 10 & 0.1 & 1 & coincident \\ 
IC 342 & 0.42 & 0.05 & 2, 3 & \nodata \\
NGC~4536 & ... & 0.01 & 4 & clumpier RC \\
NGC~6240 & $>$278 & 2.8 & 5, 6 & extra RC loop \\
NGC~6574 & 2.3 & 0.01 & 7 & Br$\gamma$ ``bar'' not seen in RC \\
NGC~6764 & 0.4 & 0.01 & 8 & outflow seen in RC and
molecular gas \\
NGC~7469 & 29 & 0.33 & 9, 10 & \nodata \\
NGC~7479 & 0.6 & ... & 11 & extra radio jet seen in RC \\
NGC~7714 & 20 & 0.15 & 1 & coincident \\
NGC~7771 & 10 & 0.06 & 12 & \nodata \\
\tableline
\end{tabular}

\tablecomments{The second column gives the circumnuclear SF rate, and the 
third column gives the SN rate, with the reference in the fourth column, if 
available. The last column gives the morphological relation of SF, as seen 
in Br$\gamma$ or H$\alpha$ lines, to radio continuum. REFERENCES.--(1) 
\citeauthor{kot01} \citeyear{kot01}; (2) \citeauthor{tur92} 
\citeyear{tur92}; (3) \citeauthor{condon82} \citeyear{condon82}; 
(4) \citeauthor{davies97} \citeyear{davies97}; (5) \citeauthor{pasq04}
\citeyear{pasq04}; (6) \citeauthor{col94} \citeyear{col94}; (7)
\citeauthor{kot00} \citeyear{kot00}; (8) \citeauthor{schin00} \citeyear{schin00}; 
(9) \citeauthor{genz95} \citeyear{genz95}; (10) \citeauthor{smith98}
\citeyear{smith98}; (11) \citeauthor{mar97} \citeyear{mar97}; (12)
\citeauthor{reu00} \citeyear{reu00}.}
\end{table*}

We also studied the morphology and location of SF, traced mostly by Br$\gamma$
emission, with respect to the radio continuum emission. The results are again
given in Table~\ref{table5}. No clear trend can be seen. While the star
formation is coincident with the radio continuum emission in a few galaxies, 
there are SF features seen in only one image, but not the other. This again
suggests that radio continuum, especially at 6 and 20 cm, is not a reliable
tracer of the location of SF activity, but it traces the overall magnetic
fields of the galaxy where electrons released in SN events spiral and produce
synchrotron radiation \citep{beck05}. Magnetic fields are only important if 
the nonthermal fraction of the emission is significant. Below we have estimated
the contribution of the thermal emission for the galaxies which had 3.5-cm
observations.

\subsection{Notes on Individual Galaxies}

\subsubsection{NGC~520}

\begin{figure*}[th]
\centering
\includegraphics[width=6in]{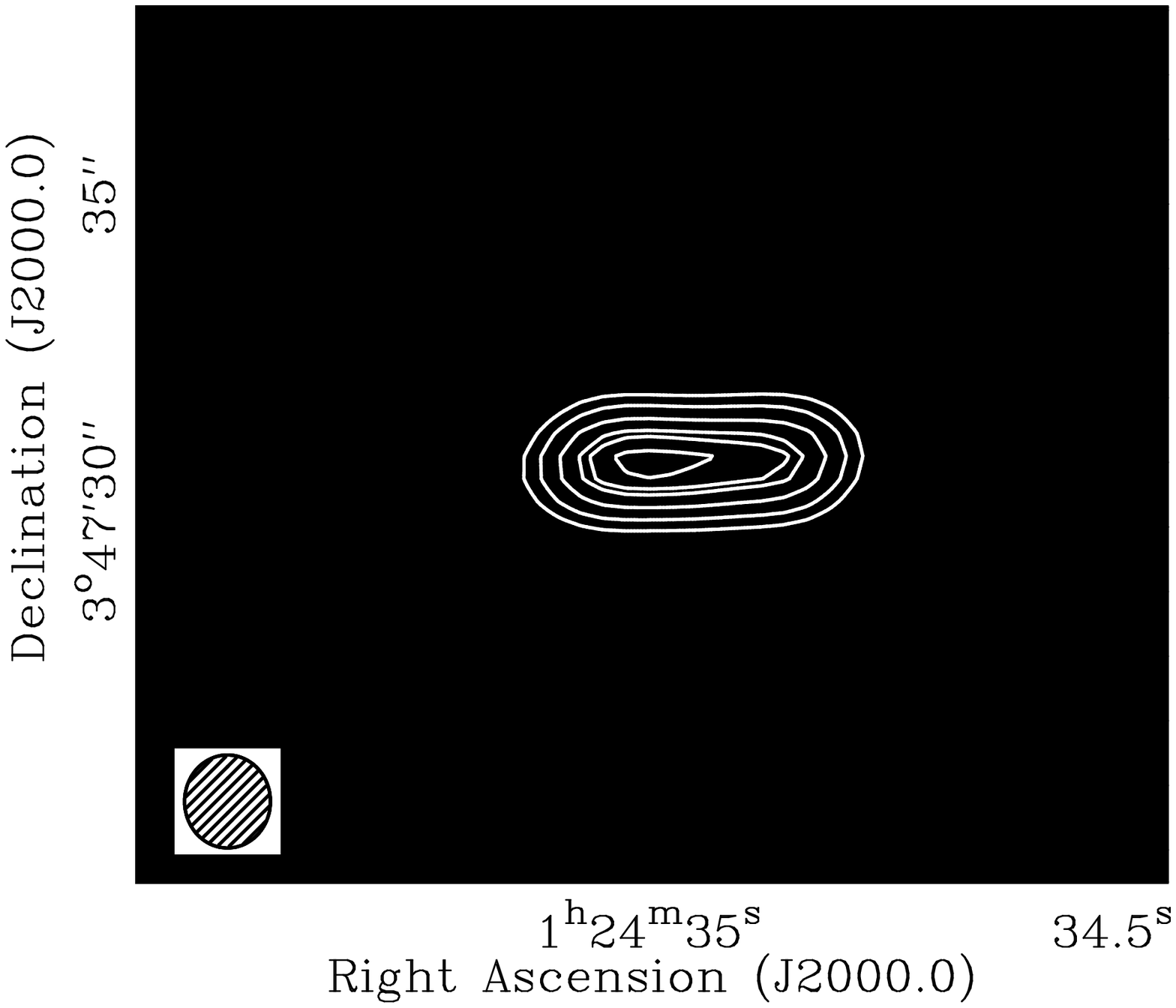}
\caption{Gray-scale image of the Br$\gamma$ emission with 20-cm radio continuum
contours overlaid in NGC~520. The contour levels are at (8 16 24 48 96 192 240 
300 356 380 420 450) times the one sigma noise level.\label{n520fig}}
\end{figure*}

NGC~520 is a merger system, presumably in an intermediate stage of merging,
according to \citet{toomre77}. The primary nucleus (the southeast nucleus) lies
behind an intricate dust lane \citep{laine03}, and the optical, including
H$\alpha$, emission is highly extinguished there \citep{bern93,laine03}.
However, the decreasing extinction in NIR allows the detection of Br$\gamma$
emission in the nuclear area \citep{kot01}. The radio continuum emission is
much more extended than the Br$\gamma$ emission, but the highest intensity,
elongated radio continuum emission region coincides well with the Br$\gamma$
emission (see Fig.~\ref{n520fig}). Since the origin of Br$\gamma$ emission is SF,
and the 20 cm radio continuum is most likely nonthermal emission from electrons
generated by supernovae in starburst regions, this spatial coincidence is
expected. The H$_{2}$ emission is slightly more extended than the Br$\gamma$
emission in NGC~520. There is no sign of circumnuclear ring structures either
in radio continuum or in optical images. This is most likely due to the fact
that the galaxy disk is highly inclined to the line of sight \citep{san88}. Our
new data are more than a  factor of two deeper than the \citet{hum87} data
taken in the same VLA configuration. The emission can be traced reliably within
a 12 arcsec $\times$ 8 arcsec area in our image, whereas the \citet{hum87} map
(only 6 cm shown, no 20 cm data shown) shows a structure of only 6 arcsec
$\times$ 2.5 arcsec in extent. Our new 20-cm image shows more extended emission
and hints at clumpy structure at lower emission levels, not seen in the 20-cm
VLA map  published by \citet{condon90}. Earlier sub-arcsec resolution VLA radio
continuum observations at 2 cm \citep*{carral90}, 6 cm \citep{condon82}, and 20
cm \citep{bes03} show a highly flattened and elongated multicomponent
structure, corresponding to the highest intensity radio continuum structure
seen in our 20 cm VLA image. The extended emission that we see in our map is
resolved out in the \citet{bes03} 20 cm map.

\subsubsection{IC 342}

\begin{figure*}[th]
\centering
\includegraphics[width=6in]{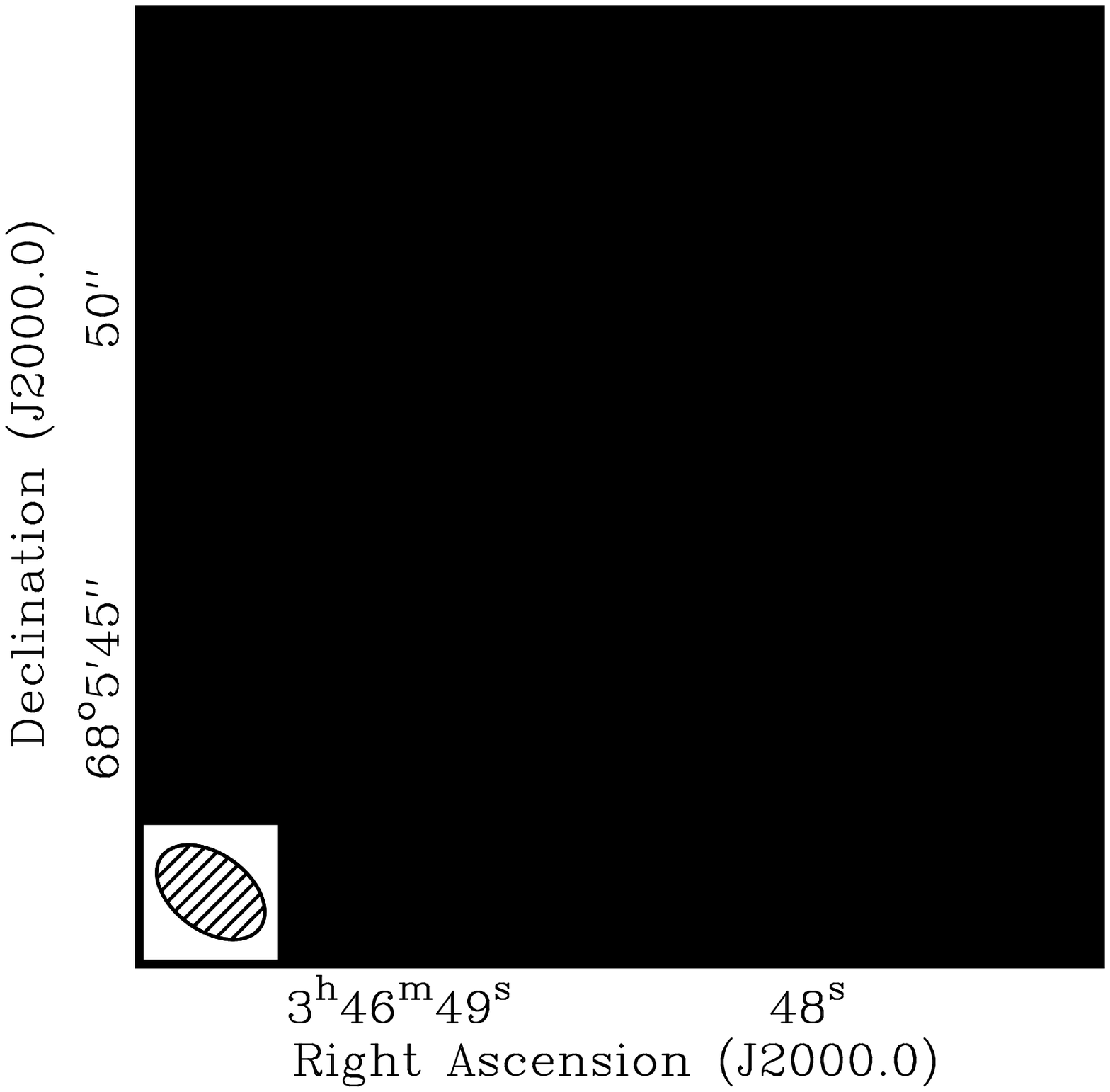}
\caption{Gray-scale image of the H$_{2}$ 1--0 S(1) emission with 20-cm radio 
continuum contours overlaid in IC~342. The contour levels are at (12 16 20 
24 32 38 48 56 64 72 88) times the one sigma noise level.\label{ic342fig}}
\end{figure*}

IC~342 is the closest actively star-forming large spiral galaxy that has an
almost face-on orientation to our line-of-sight. \citet{bauer03} suggest that
the recent starburst of at least 0.5 M$_{\odot}$~yr$^{-1}$ in IC 342 was quite
different from the archetypal starbursts, since there is a lack of hot gas and
luminous infrared emission. The nuclear cluster in IC 342 presumably formed in
a starburst about 60 Myr ago \citep*{boker97,boker99}. The registration of the
radio continuum image with the H$_{2}$ 1--0 S(1) line image
\citep[from][]{boker97} in Figure~\ref{ic342fig} is uncertain because of the
lack of absolute coordinate information in the latter image. We matched the
position of the most intense emission on the northwest side of the ring, and
checked that the northeast emission feature matches spatially. Assuming the
registration that we used is correct, the ring-like structure in radio
continuum matches well with a similar structure seen in molecular hydrogen
emission, and also in other line images shown by \citet{boker97}.
\citet{boker97} explained the ring as the inner Lindblad resonance ring of a
weak stellar bar. A ring-like structure is also seen in optical {\it HST} 
$V$-band \citep{boker99} and archival H$\alpha$ images, although the emission
maxima in optical seem to anticorrelate with the radio and near-IR emission
maps. This ring-like structure is bounded from outside by strong dust lanes,
seen in optical {\it HST} images. The overall structure in our map is similar
to the earlier  20-cm radio continuum map made at a comparable resolution by
\citet{condon82}, and the 6-cm map in \citet{turner83}. However, our maps are
more sensitive and therefore we are able to see fainter structures. Our new map
has a higher signal-to-noise ratio than the 3 cm and 6 cm observations by
\citet{beck80}.

\subsubsection{NGC~4536}

\begin{figure*}[th]
\centering
\includegraphics[width=6in]{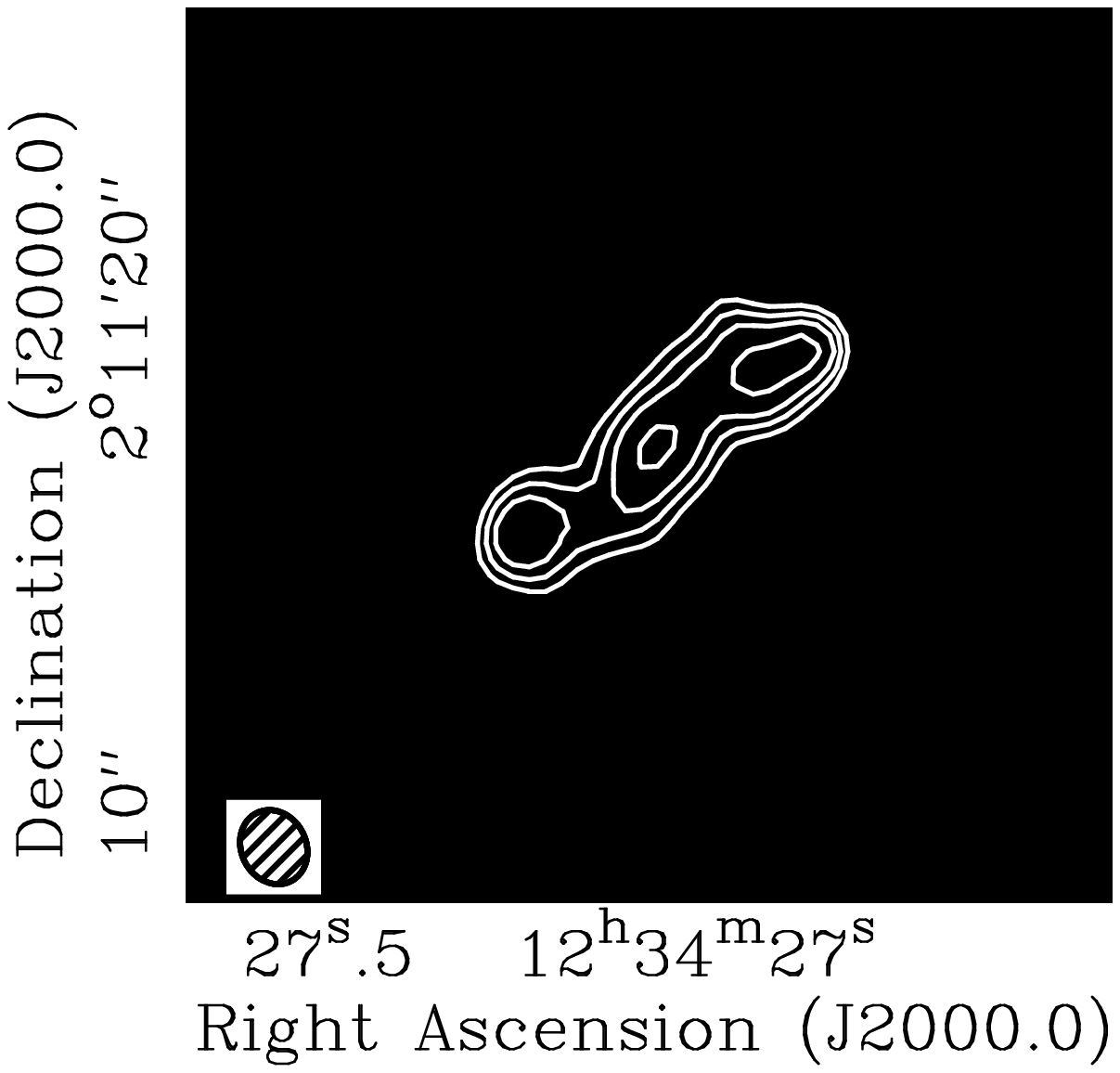}
\caption{Gray-scale image of the H$_{2}$ 1--0 S(1) emission with 20-cm radio 
continuum contours overlaid in NGC~4536. The contour levels are at (6 8 12 
16 24 32 48 64 76 84 92 104) times the one sigma noise level.\label{n4536h2}}
\end{figure*}

NGC~4536 is another nearby starburst galaxy. The molecular hydrogen 1--0 S(1)
emission near the center was investigated by \citeauthor*{davies97}
(\citeyear{davies97}; see also Fig~\ref{n4536h2}). They saw a strong 
NIR continuum peak in the nucleus, but the molecular hydrogen emission has
peaks distributed in a pseudo-ring structure around the nucleus. Archival  {\it
HST} images taken in near the $V$- and $H$-band show that the radio
continuum clumps are spatially correlated with SF regions in the
spiral arms that are close to the nucleus in projection, and not the nucleus
itself. The radio continuum observations of \citet{vila90} at 6-cm show a
similar overall  structure with peaks outside the nucleus, but their image has
a much lower sensitivity than our new data. Our new VLA data at 20 cm are at
least a factor of three deeper than the older \citet{vila90} data and show
extended  emission and three peaks lined up in the central region. Similarly,
our  observation is more sensitive than the 20-cm observation of
\citet{condon82}, allowing us to see fainter structures. 

\subsubsection{NGC~6240}

\begin{figure*}[th]
\centering
\includegraphics[width=6in]{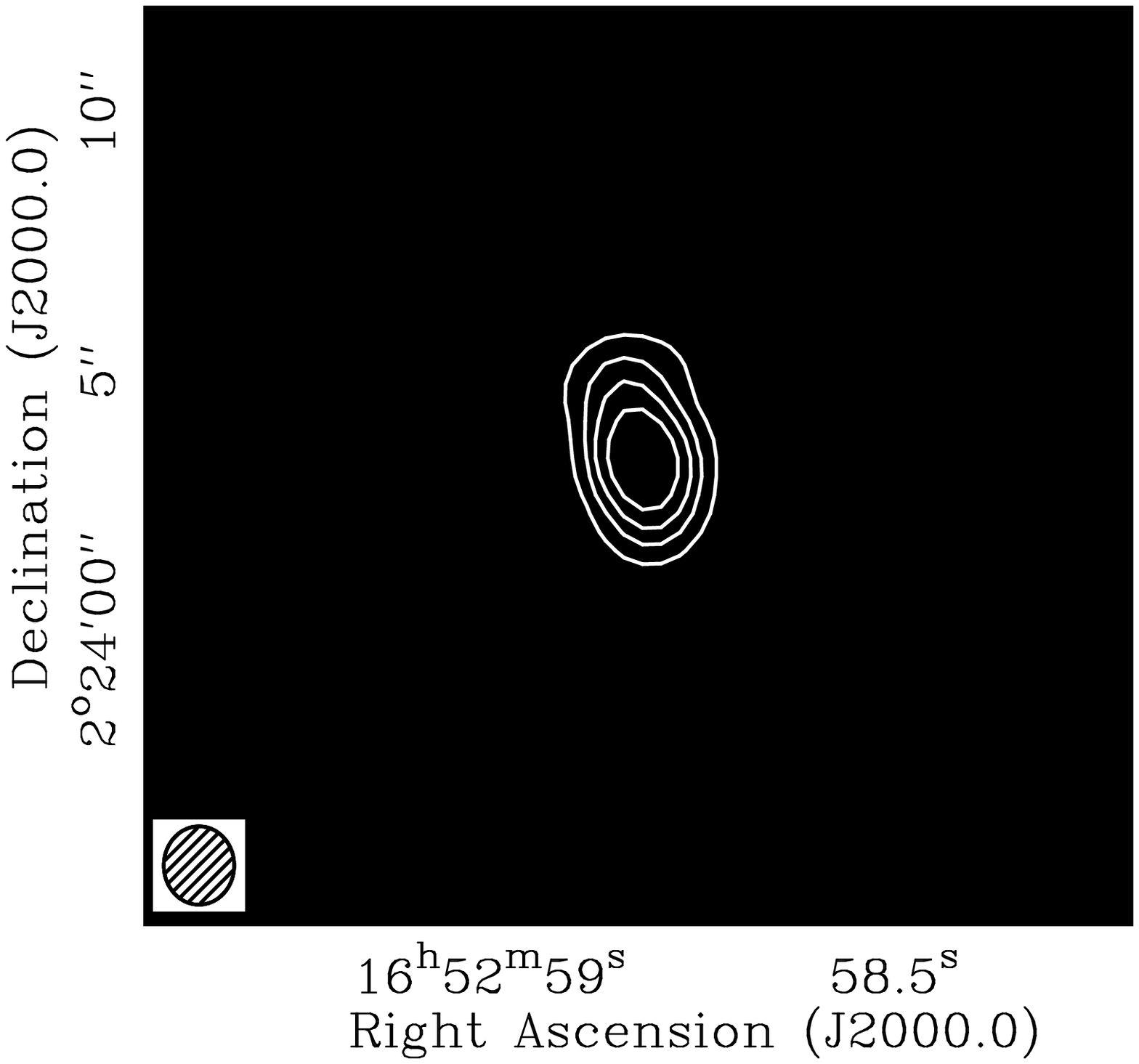}
\caption{Gray-scale image of the H$_{2}$ 1--0 S(1) emission with 20-cm radio 
continuum contours overlaid in NGC~6240. The contour levels are at (5 8 16 
24 48 96 192 300 450) times the one sigma noise level.\label{ngc6240h2}}
\end{figure*}

NGC~6240 is a famous merging system, where two nuclei are close to merging
together. The overall nuclear area of this system has LINER \citep{veilleux95}
and Seyfert 2 \citep[e.g.,][]{rafa97} classifications. Two components are seen
in sub-arcsec resolution radio continuum images
\citep[e.g.,][]{carral90,beswick01,galli04}, but our 1--2 arcsec resolution
data do not resolve the two merging nuclei, either at 3.5 cm or at 20 cm. In
fact, \citet{galli04} see even three radio components in VLBA  observations of
the nuclear region, two of which are likely to be Seyfert nuclei, based on
their radio powers and brigthness temperatures, and one which could be a radio
supernova. One of the Seyfert nuclei also has jet-like extensions. According
to \citet{lira02} the AGN component has a bolometric luminosity of about
5~$\times$~10$^{45}$~erg~s$^{-1}$, making NGC~6240 a very powerful Seyfert galaxy.
Only one peak is seen in the H$_{2}$ line observations of \citet{werf93}, who
interpret the central peak to be located between the two nuclei as a result of
a strong collision shock in the ISM (see Fig.~\ref{ngc6240h2}). The recent
H$_{2}$ 1--0 S(1) observations by \citet{max05} show a bridge of warm molecular
gas between the northern and southern active nuclei. A similar offset of the
peak from the nuclei is seen in the maps of \citet{tecza00}. The [\ion{Fe}{2}]
1.64~$\mu$m and Br$\gamma$ emission, on the other hand, coincide with the two
nuclei \citep{werf93,tecza00}. The SF activity in the nuclear
region was recently studied by \citet{pasq04}, who found extremely high SF and
SN rates in the nucleus (see Table~\ref{table5}). \citet{gerssen04} studied the
high  resolution morphology using {\it HST} NICMOS and WFPC2 observations. They
see up to four nuclear components, and a superwind from a nuclear starburst
roughly in the east--west direction. Recent Chandra observations of the nucleus
of NGC~6240 have revealed two main emission peaks, implying the existence of
two supermassive black holes, associated with the two merging nuclei
\citep{komo03}. The high SF and SN rates presumably drive a powerful
superwind.  The radio continuum emission in the nuclear region is most likely a
combination of synchrotron emission from electrons created by the high SN rate,
and non-thermal emission from the active nuclei \citep{beswick01}. Scaling the
20 cm emission (assuming it is all nonthermal and using a spectral dependence
of S $\propto$ $\nu$$^{-0.6}$, 0.6 being a typical spectral index for the
galaxies in our sample that had both 6 cm and 20 cm images) to 3.5 cm and
subtracting it we estimate that a considerable fraction of the emission at 3.5
cm can be thermal, up to 60\%. However, the spectral index of the non-thermal
emission down to 3.5 cm is not known for NGC 6240.

The most peculiar structure in this system is the extended western ``arm'' of
radio continuum emission, which has  several embedded peaks. This morphology
has been interpreted as a result of synchrotron emission from a superwind,
driven by the intense starburst in the nucleus \citep{col94}. Cosmic rays are
generated in SF regions and propagate into the magnetic features. Our
observation of the radio continuum morphology (the angle between the P.A. of
the underlying galaxy and the radio continuum emission) supports such a
scenario. It is likely that the superwind seen in detail in the
arcsec-resolution {\it HST} images of \citet{gerssen04}, is related to the
peculiar radio continuum structure. We note that NGC~6240 is the most radio
powerful galaxy in our sample, and has the highest SF and SN rates, together
with nuclear activity. Our new 3.5-cm data have a similar resolution to the
3.5-cm data discussed by \citet{col94}. We show our new data set for comparison
with the old data. \citet{anton85} published 6-cm data at a comparable
resolution. Those data reveal a similar overall morphology as our map at 20 cm,
but at a much lower sensitivity.

\subsubsection{NGC~6574}

\begin{figure*}[th]
\centering
\includegraphics[width=6in]{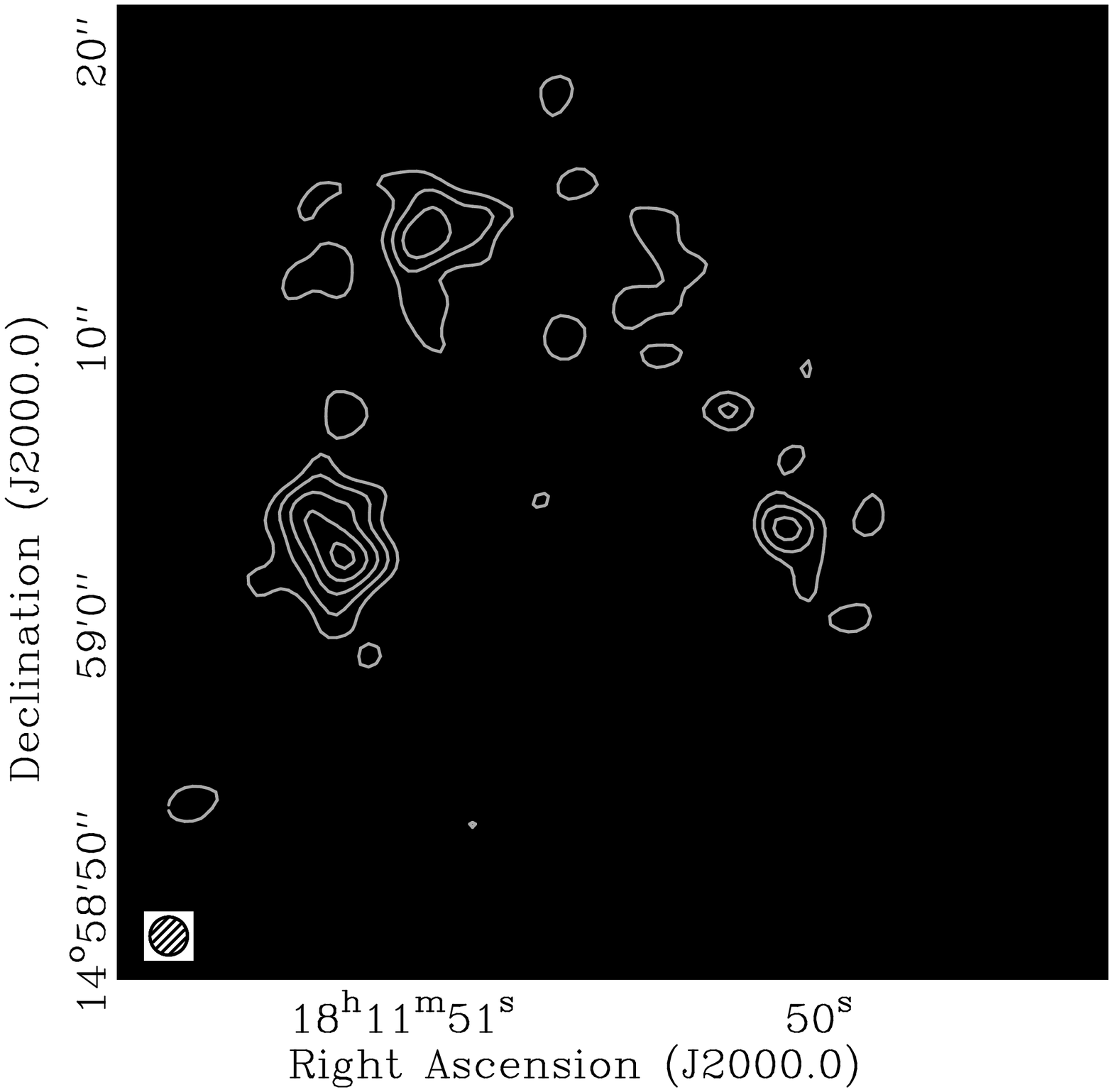}
\caption{Gray-scale image of the H$_{2}$ 1--0 S(1) emission with 20-cm radio 
continuum (black) and Br$\gamma$ (light grey) contours overlaid in NGC~6574. 
The radio continuum contour levels are at (3 4 6 8 12 16 24) times the one sigma
noise level.\label{n6574fig}}
\end{figure*}

NGC~6574 is a nearby Seyfert 2 galaxy. It is the only galaxy in our sample that
displays a central point-like radio continuum component, a circumnuclear ring,
and a connecting radio continuum feature between the two. Note that the ring
was not detected in the early VLA high resolution radio continuum observations
of \citet{vila90}, most likely because of inadequate sensitivity. Our new data
are a factor of five more sensitive than the old data published by
\citet{vila90}. The ``jet'' is one-sided and is only at the 3$\sigma$ level in
the 20-cm map (Figure~\ref{n6574fig}). At 3.5 cm, where no previous high
resolution data exist to our knowledge, there is a connection between the
nucleus and the circumnuclear ring on the opposite side of the nucleus at
6$\sigma$ level. This ``jet'' has a major-axis radius of about 10 arcsec (just
under 2 kpc). These ``jets'' lie close to the major axis of the galaxy, and
therefore are unlikely to represent any outflows perpendicular to the galaxy
disk. Interestingly, there is moderate strength H$_{2}$ emission near the end
of the southern ``jet'' (see Fig.~\ref{n6574fig}). An estimate of the thermal
emission in the 3.5 cm image, constructed in a similar way to what was
described above for NGC~6240, also shows that there is likely to be a
substantial thermal component to the 3.5 cm emission in the ring north of the
nucleus. The Br$\gamma$ emission at this location is weak, so the northeastern
radio continuum bridge is unlikely to be related to a ``jet'' striking the
circumnuclear ring and triggering of SF there. The nucleus seems to have only
nonthermal radiation at 3.5 cm, while elsewhere in the ring the thermal
contribution could be up to 50\%, based on estimating the nonthermal
contribution from the 20 cm image by scaling it down to 3.5 cm with a spectral
index of 0.6. There appears to be a nuclear stellar (and perhaps gaseous, seen
by \citeauthor{kot00} \citeyear{kot00}) bar at a position angle of about 150
degrees, inside the ring. The strongest radio continuum emission occurs in the
ring east of the nucleus, as does the strongest Br$\gamma$ and H$_{2}$ 1--0
S(1) emission (see Fig.~\ref{n6574fig}). There is a fairly good coincidence
between the H$_{2}$ 1--0 S(1) and radio continuum emission, although there is a
region of moderately strong radio continuum emission south--southeast of the
nucleus in the ring that does not show up in the near-infrared line emission.
However, the absence of any detected H$_{2}$ emission there is likely to be due
to the narrow bandwidth of the  Fabry--Perot interferometer filter used in the
observations of \citet{kot00}. An archival {\it HST} WFPC2 ultraviolet image at
3000~\AA~shows a patchy star-forming ring, but the detailed spatial correlation
with the strongest radio and NIR line emission is poor within the ring.

\subsubsection{NGC~6764}

There is a remarkable outflow structure clearly revealed by our radio
continuum images of NGC~6764, approximately perpendicular to the galaxy major
axis, and therefore probably perpendicular to the plane of the galaxy (see
Figure~\ref{fig1}). The only other earlier data at comparable resolution at 
20 cm were published by \citet{condon82}, but those data have a sensitivity
that is a factor of two worse than in our new data. The 6-cm and 21-cm
Westerbork maps by \citet{baum93} have a lower spatial resolution and
sensitivity than our maps. Our map shows clear cavities in  the radio
continuum emission, not shown by the \citet{condon82} map. Our 3.5-cm map is
the first high resolution map showing the radio continuum morphology at that
wavelength. Despite the outflow morphology, the SF and SN rates are not
remarkably high, and therefore it is possible that continuous SF models give
too low SF rates, and there actually was a starburst near the nucleus about
3--5 Myr ago, as suggested by \citet{schin00}, who also discuss the molecular
gas outflow in this galaxy. There appears to be little thermal emission in
NGC~6764 at 3.5 cm, based on the scaling of the 20 cm emission with a spectral
index of 0.6. Since this galaxy has a relatively strong stellar bar, there is
a clearly identifiable mechanism to bring fuel to the central active region.
H$_{2}$ 1--0 S(1) and Br$\gamma$ emission in NGC~6764 is relatively weak (J.
Kotilainen et al., unpublished). The nucleus of NGC~6764 has activity of
Seyfert 2 or LINER type while also having a clear starburst signature
\citep{rubin75,gonc99}. The bolometric luminosity of the nucleus in NGC~6764
is about  5~$\times$~10$^{43}$ergs~s$^{-1}$ \citep{eckart96}, including both
the  starburst and non-stellar component.

\subsubsection{NGC~7469}

NGC~7469 has a Seyfert 1.2 nucleus and a starburst ring of diameter 3 arcsec
around the nucleus \citep{miles96} is seen in a well-resolved WFPC2 image
\citep{malkan97}. In radio continuum at 1--2 arcsec resolution the nucleus
shows up as an unresolved point source which is surrounded by  diffuse emission
(see Figure~\ref{fig1}). The ring is not resolved in our radio continuum
images. \citet{condon80} shows a 6-cm image, but no 20-cm data. His 6-cm map
suffers from a highly elongated beam. \citet*{ulve81} show a 6-cm map at 1--2
arcsec resolution, but it suffers  from an elongated beam, and shows suspicious
features not seen  in our new 20-cm map. \citet{unger87} show a 20-cm map at
about 1.5--2 arcsec resolution from  observations taken in 1983. Their map has
a lower S/N than our new map, and  a poorer spatial resolution, and their map
does not show the extended  emission as well as our new map. Sub-arcsecond
resolution radio continuum images by \citet*{wilson91}, \citet{colina01},
\citet{thean01}, and \citet*{lal04} resolved the nuclear and  ring components,
but those maps do not show the extended emission seen in our new map. VLBI
observations have shown that the nucleus actually consists of three emitting
regions aligned in the east--west direction \citep{lons03}. {\it ROSAT}
observations \citep{perez96} also show centrally-peaked emission, somewhat
similar to the radio continuum morphology in our new image at 20 cm. NIR line
imaging observations show hints of the starburst ring at a radius of about 1.5
arcsec \citep{genz95}. The radio power of NGC~7469 is among the highest in our
sample, probably reflecting the powerful Seyfert 1 nucleus. The bolometric 
luminosity is about 8~$\times$~10$^{44}$~ergs~s$^{-1}$ \citep{lons03}. The star
formation rate in the nucleus, using continuous SF models, is also
very high, around 30~M$_{\odot}$yr$^{-1}$ \citep{genz95}.

\subsubsection{NGC~7479}

\citet{neff92} present a 20-cm image taken with VLA in A configuration in 1989,
and therefore, in comparable resolution to our new data. However, our new 20-cm
map has a higher sensitivity due to VLA L-band sensitivity improvement in the
early 90s by a factor of two. \citet{ho01} published  comparable resolution
maps at both 6 and 20 cm. We choose to show the map from our observations, and
it can be compared to that displayed in Figure 15 of \citet{ho01}. The jet-like
structure of NGC~7479, seen in our Figure~\ref{fig1}, was noted by
\citet{lai98}, and is the subject of a detailed study with polarized radio
continuum emission (Beck \& Laine, in preparation). The jet-like feature
probably has a magnetic origin. However, there is no clear evidence for
jet-like features that would connect to the circumnuclear region, and trigger
activity there. The jet-like structure has a total extent of several kpc, if it
lies in the galaxy plane. However, it is more likely that it is an out-of-plane
structure (Beck \& Laine, in preparation). The radio continuum has a point-like
nucleus that is unresolved even in sub-arcsec resolution MERLIN radio continuum
observations (S. Laine, unpublished). The nucleus was classified as Seyfert 1.9
by \citet{ho97b} and LINER by \citet{keel83}. The unpublished NIR line
observations show a nuclear disk (with a diameter of about 3 arcsec) of H$_{2}$
1--0 S(1) emission. The center also has a molecular gas disk seen in CO
observations \citep{lai99}, which peaks at the position of the radio continuum
peak, within the uncertainties. Archival {\it HST} NICMOS and WFPC2 images show
a strong dust lane approaching from the northwest and continuing to within a
few tens of pc from the moderately bright nucleus. No circumnuclear ring has
been detected within the strong stellar bar. The SF estimate is uncertain due
to the large amount of dust and gas near the nucleus.

\begin{figure*}[th]
\centering
\includegraphics[width=6in]{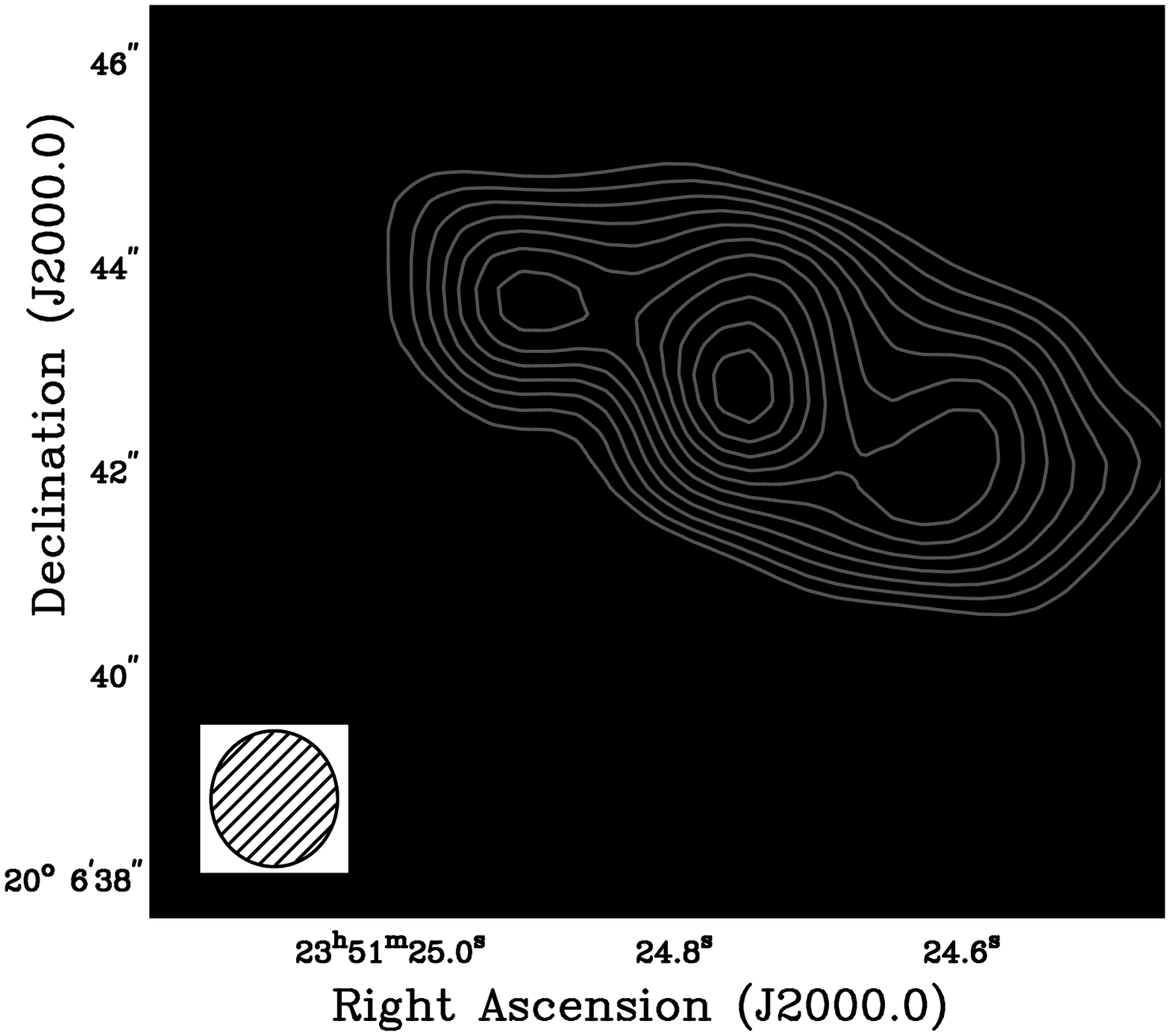}
\caption{Gray-scale image of the 6-cm radio continuum emission with 
Br$\gamma$ (black) and H$_{2}$ S=1--0 (dark grey) contours overlaid in NGC~7771.\label{n7771fig}}
\end{figure*}

\subsubsection{NGC~7714}

NGC~7714 is a nearby interacting starburst galaxy. There is a strong nuclear
and circumnuclear starburst. The Br$\gamma$ emission (and the H$_{2}$ 1--0 S(1)
emission) has an extension towards east--southeast, also hinted by our radio
continuum images at 6 and 20 cm (see Figure~\ref{fig1}). This coincides with a
spiral arm -like star forming feature seen in an {\it HST} WFPC2 F606W image
\citep*{smithb05}. In addition to the NIR line features, our radio continuum
images show a point source about three arcseconds to the west-northwest of the
nuclear component, perhaps hinted at in the H$_{2}$ 1--0 S(1) image of
\citet{kot01}.  Continuous SF models give a very high star formation rate for
NGC~7714, 20 M$_{\odot}$~yr$^{-1}$ \citep{kot01}. An earlier 6-cm image by
\citet{weedman81} has a much lower S/N, and shows suspicious structures not
seen in our new, higher sensitivity data. The 6-cm image in \citet{stine92}
does not show the same structures that are apparent in our new images, due to a
different contrast in gray-scale and contour spacing.

\subsubsection{NGC~7771}

NGC~7771 is another interacting starburst galaxy that has a clear circumnuclear
starburst ring. However, this galaxy is highly inclined (75\degr) to the line
of sight. The southwestern side of the ring has the strongest radio continuum
emission both at 6 cm and 20 cm, as well as in Br$\gamma$ emission
(Fig.~\ref{n7771fig}; NIR data from \citeauthor{reu00} \citeyear{reu00}). Our
data are a factor of five deeper than earlier 6-cm  observations at comparable
resolution by \citet*{batu92}. Sub-arcsecond resolution radio continuum
observations have resolved the ring into numerous components, and they also
reveal a nuclear radio continuum component \citep{neff92}. Note that the more
extended emission that we see in our new data are resolved out in these higher
resolution observations. The constant SF model would require very high SF
rates, and therefore it is more likely that the SF took place in a starburst
6--7 Myr ago \citep{reu00}. A very recent starburst in this galaxy was also
suggested by \citet{smith99}, who performed a case study of SF in the
circumnuclear area of this galaxy, using new near-infrared data and existing
high resolution radio images.

\section{SUMMARY}
\label{s:discussion}

Our sample of ten starburst, Seyfert, and merging or interacting galaxies has
revealed a mixed morphology in their circumnuclear regions. Ring, linear, and
``jet-like'' structures, and nuclear point sources are seen in all three types
of galaxies. We looked carefully for the connection between the nucleus and the
surrounding region, to obtain further clues about what makes galaxies active.
Our results reinforce the notion that there is no unambiguous feature
even in the radio continuum emission that would indicate how the nuclear
activity and the surrounding 1-kpc scale circumnuclear region are causally
linked to each other, or how they would evolve together. The active nuclei do
not reveal any jets or other channels by which they would trigger SF
in the surrounding disk or ring. Nor does there seem to be any particular
features in the circumnuclear rings or disks that could be used as indicators
of nuclear activity. Studies at other wavelengths have also concluded that
there are no obvious causal or spatial connections between the nuclear activity
and the properties of the circumnuclear region
\citep{storchi01,martini03,marquez04}.

We have tabulated the radio powers in the circumnuclear region of our sample
galaxies. Seyfert galaxies appear to have slightly larger average radio powers.
In galaxies that have a core component, the ratio of the core component to
circumnuclear extended radio continuum emission is typically between 30\%  and
40\%, independent of the activity or interaction class of the galaxy. The
physical size of the circumnuclear radio emission region is typically 1--2 kpc,
although there is a hint that Seyferts may have larger circumnuclear radio
emission sizes than non-Seyferts. There is no correlation between the AGN
bolometric luminosity and the radio morphology. For example, the Seyferts with
the largest AGN luminosities in our sample (NGC~6240 and NGC~7469) exhibit
different morphologies. NGC~6240 possesses small-scale jets and a nuclear
outflow, which is likely connected to the peculiar western radio continuum
loop. NGC~7469 has an aligned multicomponent radio continuum structure seen
with VLBI observations, but its circumnuclear radio morphology seen in our 1.4
arcsec resolution observations  does not present any specific structures, only
a nuclear component with some diffuse emission around it. The AGN with lower
bolometric luminosities have a widely varying radio continuum structure as
well. NGC~6574 has a circumnuclear radio continuum ring, NGC~6764 has an
outflow structure in radio continuum, and finally NGC~7479 has peculiar
large-scale radio ``jets.'' A nuclear starburst  exhibits itself in various
ways in the radio continuum morphology. NGC~6764 has a clear outflow
morphology, while NGC~6240 has a peculiar loop structure on one side of the
nucleus, possibly as a result of a superwind from several supernova explosions
in the nucleus \citep{col94}. Perhaps the most intriguing structure  is seen in
NGC~7479, which has a jet-like morphology, together with a core component, but
no extended nuclear starburst region. This galaxy has been speculated to be
experiencing a minor merger \citep{laine99b}. It also has a massive molecular
disk around the nucleus \citep{lai99}. Future research is warranted into
investigating any connection between a minor merger, the nuclear activity, and
the radio ``jet.''

We have compared the radial surface brightness profiles of radio emission and
Br$\gamma$ and/or H$_{2}$ line emission. While the starburst rings show up
clearly in the Br$\gamma$ emission, we see no systematic differences between
the Seyfert and starburst galaxies in these profiles. We also calculated the
CAS parameters for our sample galaxies. Again, the average concentration,
asymmetry, and clumpiness parameters are not significantly different between
Seyfert and starburst galaxies. The orientation of the radio continuum emission
differs from the orientation of the galaxy major axis more in Seyferts than in
starbursts. We have also used SF and SN rates found in the literature to
estimate how they depend on the distribution of radio continuum emission and
the activity class of the galaxy. We have inspected the correlation of the SF
and SN rates with the strength non-stellar nuclear activity or interaction
class, but found no clear trends.

Finally, we have compared the radio continuum morphology to that of NIR line
emission morphologies, including Br$\gamma$ and H$_{2}$ line emission, when
available. We find that the bulk of the radio continuum usually traces the sum
of the Br$\gamma$ and H$_{2}$ line emission fairly reliably. However, several
extranuclear radio continuum emission features not visible in the NIR line
images were found, including the jet-like structures and the outflow-type
structures. These differences are likely caused by the effects of dust
absorption and scattering (affecting near-IR emission) and the spatial 
distribution and strength of magnetic fields (affecting radio continuum
emission).

Our results will be extended and validated in a follow-up study that includes
a much larger sample of Seyfert and starburst galaxies. The  information
content of the radio continuum observations will be increased by including
polarization information. Unfortunately, such observations are time-expensive,
and the next step we intend to take includes a detailed study of the magnetic
origin of the jet-like structure in NGC~7479.


\acknowledgments 

We thank Torsten B\"{o}ker, Richard Davies, and Paul van der Werf for providing
us with near-infrared line images. We are also thankful to Ray Norris for
comments on an earlier draft of this paper, to Mark Lacy for discussions on
radio jet--AGN luminosity correlation, and to Tom Pannuti for discussions on the
radio continuum emission from supernova remnants. We are thankful to the
referee, W. C. Keel, for suggestions that improved the quality of the paper.
JKK was supported by the Academy of Finland (project 8201017) during part of
this work. The National Radio Astronomy Observatory is a facility of the
National Science Foundation operated under cooperative agreement by Associated
Universities, Inc. This research has made use of the NASA/IPAC Extragalactic
Database (NED) which is operated by the Jet Propulsion Laboratory, California
Institute of Technology, under contract with the National Aeronautics and Space
Administration. This publication makes use of data products from the Two Micron
All Sky Survey, which is a joint project of the University of Massachusetts and
the Infrared Processing and Analysis Center/California Institute of Technology,
funded by the National Aeronautics and Space Administration and the National
Science Foundation.

\end{document}